\input{aipcheck}

\documentclass[
    ,final            
  ]
  {aipproc}

\layoutstyle{8x11single}

\begin{document}

\title{The search for progenitor models of type Ia supernovae}


\author{J.S.W. Claeys}{
  address={Sterrekundig Insituut, Universiteit Utrecht, PO Box 800000, 3508 TA Utrecht, The Netherlands}
}

\author{O.R. Pols}{
  address={Sterrekundig Insituut, Universiteit Utrecht, PO Box 800000, 3508 TA Utrecht, The Netherlands}
}

\author{J. Vink}{
  address={Sterrekundig Insituut, Universiteit Utrecht, PO Box 800000, 3508 TA Utrecht, The Netherlands}
}

\author{R.G. Izzard}{
  address={Argelander Institute for Astronomy, University of Bonn, Auf dem Huegel 71, D-53121 Bonn, Germany} 
}

\begin{abstract}
We show the preliminary results of our search for the progenitor systems of type Ia supernovae (SNe Ia). We model binary populations our aim being to compare these models with the observations of detailed element abundances of the hot Intra-Cluster Medium.
\end{abstract}

\maketitle


\section{Introduction}

Type Ia supernovae (SNe Ia) are thermonuclear explosions of carbon-oxygen white dwarfs (CO WDs). They are used as cosmological distance indicators \citep{Perlmutter99,Riess96, Riess98, Phillips93} and are the main source of iron in the Universe. 
However, the nature of their progenitors is not well understood. 
Two progenitor scenarios have been proposed, namely the single degenerate (SD) and double degenerate (DD) scenario. In the DD scenario, 2 CO WDs are formed close together by previous common envelope phases and merge due to gravitational radiation. In the SD scenario a CO WD accretes material from a companion that loses material by Roche lobe overflow (RLOF) or by a stellar wind. In this research we distinguish two types of companions, namely a hydrogen rich companion (SD$_{\rm H}$-scenario) and a more evolved helium rich companion (SD$_{\rm He}$-scenario).
\newline The observed delay time distribution (DTD) of SNe Ia, which is the rate of supernovae as a function of time since formation, can be used to constrain the progenitor models. In addition, the nucleosynthetic-yields of SNe Ia can be used to test the progenitor models. Clusters of galaxies are gravitationally bound systems in which metals from SNe, stellar winds, etc., accumulate over a long time span. They are ideal tools to examine the time integrated yields of SNe Ia. In this research we will study the evolution of SNe Ia using a binary population synthesis code based on single and binary evolution models \citep{Hurley00, Hurley02} in combination with a synthetic nucleosynthesis model \citep{Izzard04, Izzard06, Izzard09}. We expect that the combination of the DTD, the chemical evolution of binaries and the observations of clusters will give us new insights into the evolution of stars towards type Ia SNe.

\section{Binary population synthesis}
\begin{figure}
  \includegraphics[height=.30\textheight]{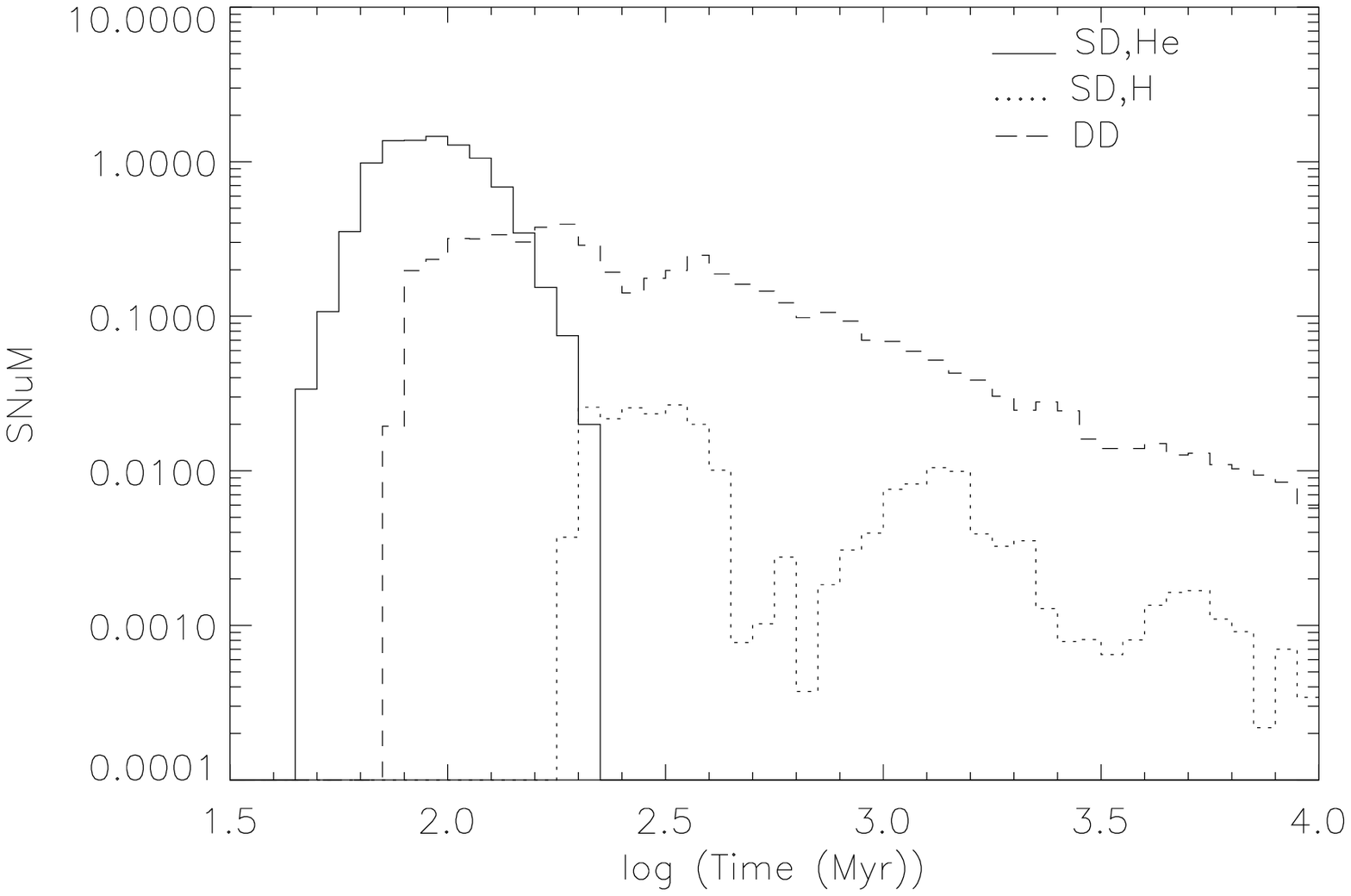}
  \includegraphics[height=.30\textheight]{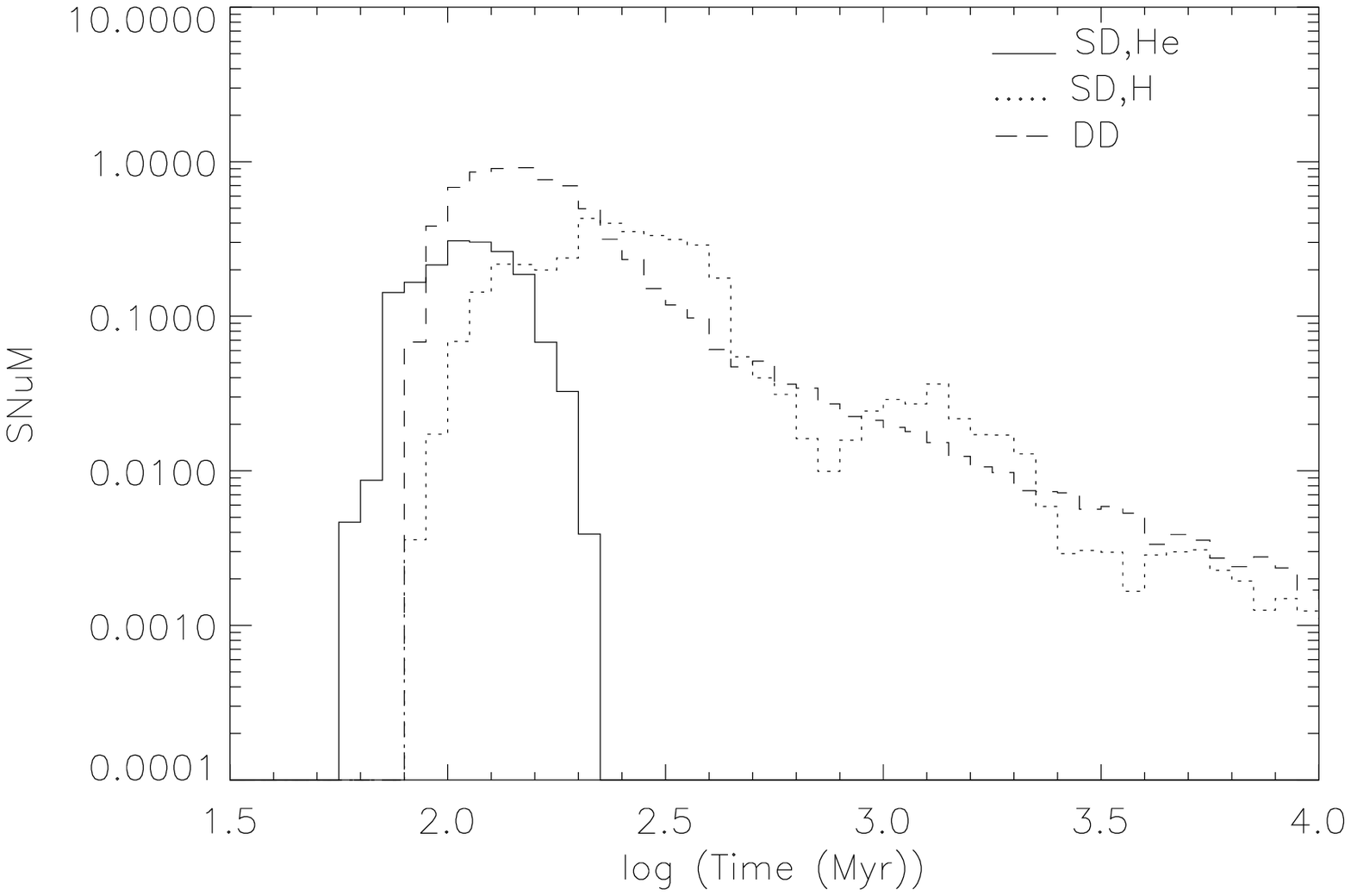}
  \caption{The Delay Time Distribution (DTD) of SNe Ia in SNuM of the three different progenitor scenarios in our model (Left Fig. $\alpha_{\rm CE}$ = 3. Right Fig. $\alpha_{\rm CE}$ = 1). The solid line indicates the rate of the SD$_{\rm He}$ channel, the dotted line the SD$_{\rm H}$ channel and the dashed line the DD channel.}
  \label{DTD}
\end{figure}

As a first step in this project we simulate binary populations in order to compare the resulting DTD to observations as well as to the results of other studies. The observed DTD shows a prompt peak \citep{Aubourg08} and a delayed component, which follows a power law, namely $t^{-1}$ \cite{Totani08}.
We incorporate the efficiency of surface H and He burning on a WD and the model for an optically thick WD wind, based on \citet{Hachisu99} \& \citet{Kato04}, into our binary population synthesis code. Other ingredients of our code and the initial distributions of binary parameters are as described by \citep{Hurley00, Hurley02}.
In Fig. \ref{DTD} we show the preliminary results of two simulations, in which we vary the common envelope (CE) efficiency; $\alpha_{\rm CE}$ = 3 (left panel) and $\alpha_{\rm CE}$ = 1 (right panel). The lower value of $\alpha_{\rm CE}$ results in closer binary orbits after the CE phase \citep{Hurley02}. The simulations show that the three different progenitor channels contribute to the total rate of SNe Ia. Both our simulations show that the SD$_{\rm He}$ channel is dominant between 50 and 200 Myr, with an average rate over this time interval of 0.1-0.6 SNuM \footnote{SNuM = Supernova rate per 100 yr per 10$^{10}$ M$_\odot$ in stars}, depending on the common envelope prescription. 
The dominant scenario between 200 Myr and a Hubble time is the DD channel, with an average rate of 0.01-0.02 SNuM. 
The minor channel is SD$_{\rm H}$, with a rate about one magnitude lower then the DD channel in the case of a high $\alpha_{\rm CE}$. The average rate of type Ia SNe over a Hubble time is about 0.03 SNuM in both cases.
Our results reproduce, at least qualitatively, the prompt peak and delayed $t^{-1}$ dependence of the observed DTD. But our simulated rate is a factor of 2-3 lower than the observed one \cite{Totani08}.
\newline Other population synthesis studies also find a prompt peak caused by the SD$_{\rm He}$ channel \cite[e.g.][]{Wang10} and the power law dependence of the delayed SNe Ia \cite[e.g.][]{Ruiter09}. However, the largest deviation between the  different models is in the SD$_{\rm H}$ channel, due to the uncertainties in the accretion rate and accretion efficiency of H and He bunring on a WD and the efficiency of wind accretion onto a WD \citep{Hachisu99}. For example, \citep{Ruiter09} find $\sim$ 10$^{-3}$ SNuM for the SD$_{\rm H}$ channel, while \citep{Wang10a} find a rate which is about an order of magnitude higher.
\newline With this research we hope to constrain the DTD by means of binary population synthesis and to eliminate the uncertainties in single and binary evolution.

\bibliographystyle{aipproc}   

\bibliography{lit}

\IfFileExists{\jobname.bbl}{}
 {\typeout{}
  \typeout{******************************************}
  \typeout{** Please run "bibtex \jobname" to optain}
  \typeout{** the bibliography and then re-run LaTeX}
  \typeout{** twice to fix the references!}
  \typeout{******************************************}
  \typeout{}
 }

\end{document}